\documentclass [aps,prb,twocolumn] {revtex4}
\usepackage[final]{graphics}
\usepackage{amssymb}
\usepackage{amsfonts}
\usepackage{epsfig}

\begin{document}

\date{\today}

\title{Critical dynamics of the simple-cubic Heisenberg antiferromagnet RbMnF$_3$: Extrapolation to q=0}

\author{Shan-Ho Tsai$^{a,b}$ and D. P. Landau$^a$}
\affiliation{
$^a$ Center for Simulational Physics, University of Georgia, Athens, GA 30602\\
$^b$ Enterprise Information Technology Services, University of Georgia, Athens, GA 30602}

\begin{abstract}
Monte Carlo and spin dynamics simulations have been used to study
the dynamic critical behavior of RbMnF$_3$, treated as a classical 
Heisenberg antiferromagnet on a simple cubic lattice. In an attempt to
understand the difference in the value of the dynamic critical
exponent $z$ between experiment and theory, we have used larger lattice sizes
than in our previous simulations to better probe the asymptotic
critical region in momentum. We estimate $z=1.49\pm 0.03$, in good 
agreement with the renormalization-group theory and dynamic scaling 
predictions. In addition, the central peak in the dynamic structure factor
at $T_c$, seen in experiments and previous simulations, but absent in
the renomalization-group and mode-coupling theories, is shown to be 
solely in the longitudinal component. 
\end{abstract}

\pacs{75.10.Hk,75.40.Gb,75.40.Mg,64.60.Ht}

\maketitle

\section{Introduction}

The dynamic critical behavior of RbMnF$_3$, a close realization of the
isotropic, simple cubic Heisenberg antiferromagnet, has been the subject
of several experimental (see Coldea et al\cite{coldea} and references therein),
and theoretical studies\cite{mazenkoTc,mazenkoTb,cuccoli,hohenhal,halhoh69}. 
The real time dynamics of this system is governed by coupled equations of 
motion for the magnetic ions. In the classification of Hohenberg and 
Halperin\cite{hohenhal}, the critical dynamics of this system pertains 
to class G, for which the order parameter (the staggered magnetization) 
is not conserved. 

Dynamic critical behavior can be characterized by a dynamic critical 
exponent $z$, and for RbMnF$_3$ the most precise experimental 
estimate\cite{coldea} is $z=1.43\pm 0.04$. This estimate is slightly below
the predicted value\cite{mazenkoTc,cuccoli,hohenhal,halhoh69} of $z=1.5$ 
for an isotropic three-dimensional Heisenberg antiferromagnet. Our 
previous estimate using spin-dynamics simulation\cite{sd1} 
(obtained before we knew the latest experimental result\cite{coldea})
was $z=1.43\pm 0.03$ in good agreement with the experimental value.
The slight disagreement between theory and both the experiment and 
the simulation was perplexing, but one possible explanation was that 
neither of these latter studies probed the asymptotic critical region 
sufficiently far.
To check this, neutron scattering 
experiments would have to be performed with smaller momentum transfer
$q$, a task that can be quite challenging. We can also resort to spin dynamics
simulations of the model on larger lattice sizes which allow access to 
smaller $q$ values.

Spin dynamics simulations\cite{DPLMKreview} have proven to be an 
effective tool to study dynamic behavior of magnetic systems.
Direct and quantitative comparisons of magnetic excitation dispersion 
curves and dynamic structure factor line shapes from experiment\cite{coldea} 
and spin dynamics simulations have shown good agreement, with no adjustable
parameters\cite{sd1,sd2}.
At the critical temperature $T_c$, both experiment and simulation find that
the average dynamic structure factor $S({\bf q},\omega)$, for momentum 
${\bf q}$ and frequency $\omega$, has a 
spin-wave peak and an additional central peak not predicted by either
the renormalization-group theory\cite{mazenkoTc} or mode-coupling 
theory\cite{cuccoli}. In a polarized neutron scattering 
experiment\cite{cox} below
$T_c$, the quasi-elastic peak has been shown to be longitudinal in character.
It is thus interesting to investigate whether the
central peak at $T_c$ originates in the longitudinal, in the transverse 
or in both components with respect to the staggered magnetization. 
Another motivation for separating the longitudinal and transverse components
of the dynamic structure factor is to test the theoretical 
prediction\cite{mazenkoTb} that below $T_c$ 
the longitudinal momentum-dependent susceptibility behaves as 
$\chi^L(q)\sim 1/q$, whereas the $q$-dependence of the 
transverse component is $\chi^T(q)\sim 1/q^2$. Although experimental 
measurements\cite{coldea} are consistent with $\chi^L(q)\sim 1/q$, 
the lack of reliable 
small wave vector measurements hindered a conclusive experimental test of 
this predicted divergence. The experimental value for the transverse component
is $\chi^T(q)\sim 1/q^{1.91\pm 0.05}$.

In this paper we use spin dynamics simulations to study the dynamic
critical behavior of the isotropic Heisenberg antiferromagnet on the
simple cubic lattice, using larger lattices than in our previous simulations,
as motivated above. We investigate the longitudinal and transverse components 
of the dynamic structure factor, and compare the $q$-dependence of the 
susceptibility with the predicted forms both at and below $T_c$.
For brevity, we do not present our methodology in great detail here; it
has been described in earlier work\cite{sd1,DPLMKreview}.

\section{Model and Methods}

We consider three-dimensional classical spins ${\bf S_r}$ of unit length,
defined on the sites ${\bf r}$ of $L\times L\times L$ simple cubic lattices.
The interaction is described by a model Hamiltonian written as
\begin{equation}
{\cal H}=J\sum_{<{\bf rr^{\prime }}>}{\bf S_{r}}\cdot {\bf S_{r^{\prime }}},
\end{equation}
where the summation is over pairs of nearest-neighbor
spins, and $J>0$ is an antiferromagnetic exchange coupling. 
An earlier high-resolution Monte Carlo simulation\cite{kunTc} 
determined $T_c=1.442929(77)J$, so any uncertainty in the location of the
critical temperature of this model is negligible.

The dynamics of the spins is governed by coupled equations of 
motion\cite{DPLMKreview} and the dynamic structure factor 
$S^k({\bf q},\omega )$ is given by the Fourier transform 
of the space- and time-displaced correlation functions 
$C^k({\bf r- r^{\prime}},t) =\langle S_{\bf r}^k(t)
S_{\bf r^{\prime}}^k(0)\rangle- \langle S_{\bf r}^k(t)\rangle\langle 
S_{\bf r^{\prime}}^k(0)\rangle$,
where $k$ represents the spin components.
The coupled equations of motion were integrated 
using an algorithm\cite{krech} based on 4th-order Suzuki-Trotter 
decompositions of exponential operators, with a time step $dt$. 
The integrations were performed up to time $t_{max}$, 
starting from equilibrium configurations at temperature $T$  
generated with a hybrid Monte Carlo method\cite{sd1}. 
The correlations $C^k({\bf r- r^{\prime}},t)$ were computed for time 
displacements ranging from $0$ to $t_{cutoff}$ and the canonical ensemble
was established by averaging results from $N$ different initial
equilibrium configurations. 
We used periodic boundary conditions in space and thus we could only access
a set of discrete values of momentum transfer given by
$q=2\pi n_q/L$, where $n_q=1, 2,...,L/2$. Because of computer memory
limitations we restricted our data to ${\bf q}=(q,0,0)$, $(q,q,0)$, 
and $(q,q,q)$, which correspond to the [100], [110], and [111] directions, 
respectively.

In an attempt to probe the true asymptotic critical region in momentum and
thus provide a more accurate estimate of the dynamic critical exponent, 
we extended our Monte Carlo\cite{MCbook} and spin-dynamics 
simulations\cite{DPLMKreview,sd1} to systems as large as $L=72$ at $T_c$. 
For $L=72$, each equilibrium configuration was generated with 2500 hybrid 
Monte Carlo steps, each of which consists of two Metropolis sweeps through
the lattice and eight overrelaxation steps. We used $dt=0.2/J$, 
$t_{max}=1080/J$, $t_{cutoff}=1000/J$, and $N=1000$. We have also
improved the statistics of our previous simulations for $L=48$ and $60$
by increasing the number of initial configurations to $N=1000$.
With $L=72$, the smallest wave vector that we can access is 
$q=2\pi/72=2\pi(0.0139)$, whereas, in our units, the smallest $q$ 
value probed by experiment\cite{coldea} was $q=2\pi(0.02)$.

The dynamic critical exponent $z$ can be extracted using dynamic  
finite-size scaling\cite{DPLMKreview}, which yields 
$\omega\bar S^k({\bf q},\omega)/{\bar{\chi}}^k({\bf q})=
G(\omega L^z,qL,\delta_{\omega}L^z)$ and 
$\bar\omega_m^k=L^{-z}\bar\Omega^k(qL,\delta_{\omega}L^z)$,
where $k$ represents the polarization, $\bar {S}^k({\bf q},\omega)$ 
is the dynamic structure factor
convoluted with a Gaussian resolution function with parameter 
$\delta_{\omega}$,
${\bar{\chi}}^k({\bf q})$ is the total integrated intensity, and
$\bar\omega_m^k$ is a characteristic frequency, defined by 
$\int_{-\bar\omega_m^k}^{\bar\omega_m^k}\bar S^k({\bf q},\omega)d\omega/
2\pi ={\bar{\chi}}^k({\bf q})/2$. If we set $\delta_{\omega}=0$ the exponent
$z$ can be obtained from the slope of a graph of ln$(\omega_m^k)$ vs ln$(L)$,
at fixed $qL$. To test the robustness of the estimate, we have also used
$\delta_{\omega}=0.005(72/L)^z$, and determined $z$ 
iteratively\cite{DPLMKreview,sd1}.
As in our previous work\cite{sd1,sd2}, the dynamic critical exponent
is estimated using the average dynamic structure factor.

After understanding the difference in the dynamic critical exponent
between experiment and theory, we can examine other
unresolved issues, such as the nature of the central peak at $T_c$ 
observed in experiments, but absent in theoretical studies. To this
end, we analyzed longitudinal and transverse motions of the 
spins with respect to the staggered magnetization. 
The dynamics of the isotropic Heisenberg model, given by the coupled
equations of motion, conserves both the total energy and the
uniform magnetization, which is the order parameter for the Heisenberg
ferromagnet. However, for the antiferromagnet the order parameter 
(staggered magnetization) is not a constant of the motion; therefore,
separating components of the spin parallel (longitudinal
component) and perpendicular (transverse component) to the order parameter
is challenging. Our approach to determine the individual components
of the spin motion, and thus of $S({\bf q},\omega)$, 
was to rotate the frame of reference to align the $z$-axis
parallel to the staggered magnetization before starting the time 
integrations, to make the $z$-axis coincide with the longitudinal
direction. As we integrated the equations of motion, the direction of the
staggered magnetization changed slightly because it is not a conserved 
quantity. Therefore, after each integration step we re-rotated the frame
of reference to re-align the $z$-axis with the staggered magnetization,
thereby restoring the $z$-axis as the longitudinal direction.
In this part of our simulations we used $L=24$ 
with $t_{max}=480/J$, $t_{cutoff}=400/J$, $dt=0.2/J$, and $N=12,000$,
in addition to $L=36$, $48$, and $60$, with $t_{max}=880/J$,
$t_{cutoff}=800/J$, $dt=0.2/J$, and $N=11,000$, $7,000$, and $3,000$, 
respectively. 
We denote the longitudinal and transverse components of 
$S({\bf q},\omega)$ as $S^L({\bf q},\omega)$ and $S^T({\bf q},\omega)$, 
respectively. 

We have also investigated ${\chi}^L(q)$ and ${\chi}^T(q)$, the integrated 
intensities of $S^L({\bf q},\omega)$ and $S^T({\bf q},\omega)$, respectively.
When the decay of the dynamic structure factor is slow, as is the case
of $S^T({\bf q},\omega)$ at $T_c$, computing the integrated intensity 
is complicated by the fact that at high frequencies 
[$\pi/(2dt)\lesssim \omega \lesssim \pi/dt$] the approximation 
of the integral Fourier transform as a discrete sum is highly dependent 
on the summation method 
(e.g. direct sum, trapezoidal rule, Simpson's rule, etc.). 
In terms of cpu time usage, it is more efficient to integrate 
the equations of motion with the largest time step $dt$ that still yields
a stable method and accurate time evolutions; however, the maximum accessible 
frequency is inversely proportional to $dt$. 
Simulations for $L=24$ at $T_c$ using a smaller time step ($dt=0.1/J$) 
showed that both the Simpson's rule and the trapezoidal rule  
produce consistent results for $S^L({\bf q},\omega)$ and 
$S^T({\bf q},\omega)$ up to $\omega\approx 15J$.
A direct comparison between the dynamic structure factor obtained with
 $dt=0.1/J$ and that
obtained with the Simpson's rule and the trapezoidal rule using $dt=0.2/J$
up to $\omega\approx 15J$ shows that the trapezoidal rule gives a better 
approximation for $S^L({\bf q},\omega)$ and
$S^T({\bf q},\omega)$ at high frequencies. Estimates of ${\chi}^L(q)$ 
and ${\chi}^T(q)$ are not significantly dependent on the methods used to
integrate $S^L({\bf q},\omega)$ and $S^T({\bf q},\omega)$, respectively,
and we use Simpson's rule for these integrations.

Our approach to determine ${\chi}^T(q)$ at $T_c$ was to compute
$S^T({\bf q},\omega)$ with the trapezoidal rule and then to integrate it 
from $\omega=0$ to $8J$.  
Although the intensity of $S^T({\bf q},\omega)$ has not 
decayed to zero at $\omega = 8J$, it is a small fraction of
the intensity of the spin-wave peak. 
Data for $\omega=4J$ to $8J$ were fitted with an exponential function, which
was then used in the integration to $\omega=\infty$.
(An exponential function was chosen
for this fitting because it is the simplest one that yielded a good 
fitting in the range of frequency used.)  
To estimate the longitudinal component ${\chi}^L(q)$ at $T_c$ we simply 
used the trapezoidal rule to determine $S^L({\bf q},\omega)$ and then 
integrated it from $\omega=0$ to $\omega=10J$, without further 
high-frequency corrections, because $S^L({\bf q},\omega)$ decays quickly in
frequency. 

The behavior of ${\chi}^T(q)$ and ${\chi}^L(q)$ below $T_c$ is studied with
spin dynamics simulations at $T=0.5T_c$ using $dt=0.2/J$ and 
$L=24$ and $36$, with $N=1000$ for each lattice size.
At this temperature both $S^L({\bf q},\omega)$ and $S^T({\bf q},\omega)$
decay quickly, hence these line shapes were obtained with the trapezoidal rule
and they were then integrated from $\omega=0$ to $\omega=10J$.

For comparison, we have also computed the longitudinal and transverse 
components of $S({\bf q},\omega)$ with respect to the residual uniform 
magnetization. In this case, the rotation of the frame of reference
to align one axis parallel to the uniform magnetization was only
required once, before starting the time integrations, because the uniform
magnetization vector is conserved. 

\section{Results}

\begin{figure}[ht]
\includegraphics[clip,angle=0,width=8cm]{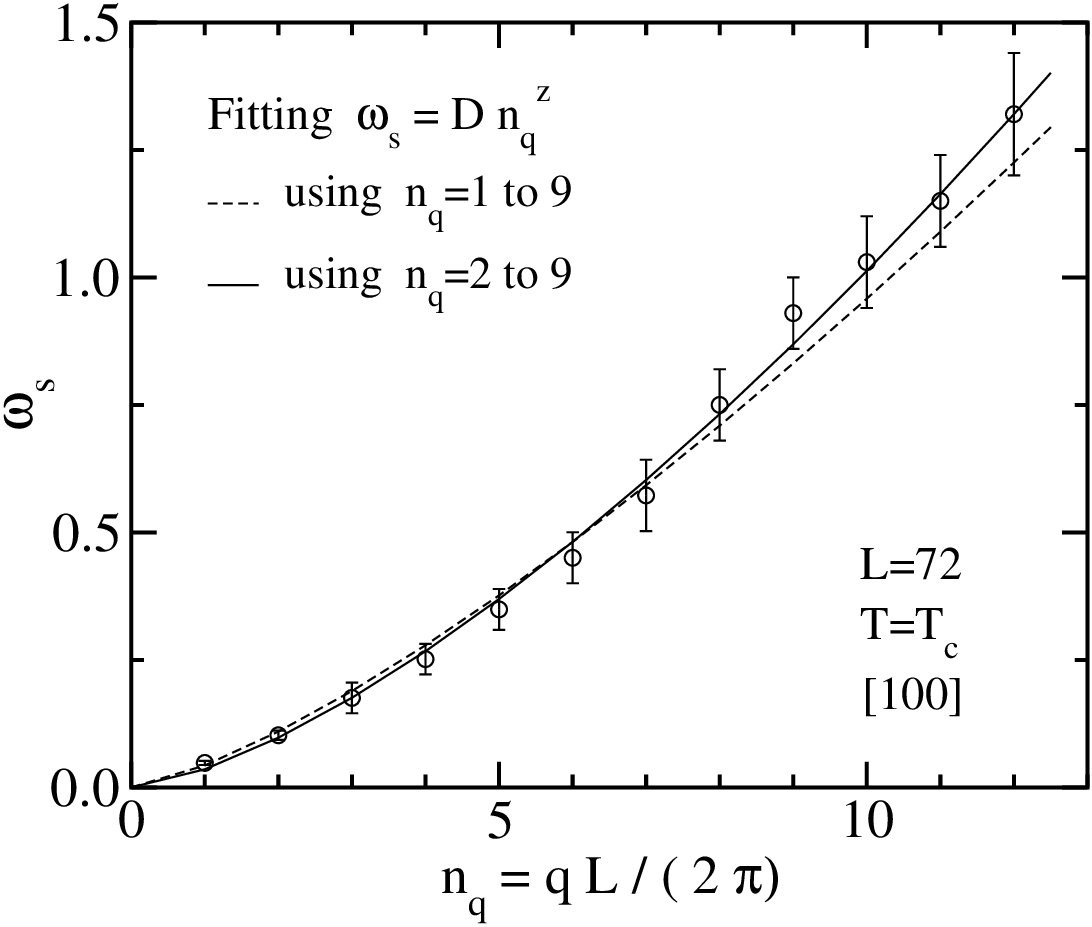}
\caption{\label{figwq}Spin-wave dispersion curve for the average 
dynamic structure factor.}
\end{figure}

Our results for the average dynamic structure factor $S({\bf q},\omega)$
show a spin wave and a central peak for $T\le T_c$. The spin wave dispersion
curve for $L=72$, $T=T_c$ and ${\bf q}$ vectors in the [100] direction, 
shown in Fig.\ref{figwq}, was obtained by fitting $S({\bf q},\omega)$ with
Lorentzian spin-wave creation and annihilation peaks centered at 
$\omega=\pm\omega_s$, and a Lorentzian peak at $\omega=0$. 
We have obtained
reasonable fittings for $q$ values corresponding to $n_q=1$ to $7$; for 
larger $n_q$, the spin-wave line shapes were not Lorentzian and the 
positions of the spin-wave peaks $\omega_s$ were read off directly 
without any fitting. We remind the reader that the first Brillouin zone edge
in the [100] direction occurs at $n_q=L/2$.
An estimate of the dynamic critical exponent $z$ is obtained by fitting
the dispersion curve with a function\cite{hohenhal} $\omega_s=Dn_q^z$. 
Since the asymptotic
critical region corresponds to small values of $q$, we have used $n_q=1$ to
$9$ in the fitting (dashed line in Fig.\ref{figwq}), and obtained 
$z=1.35\pm 0.05$. If $n_q=1$ is excluded due to its large finite-size 
effect\cite{sd1}, the fitting yields 
$z=1.45\pm 0.07$ (solid line in Fig.\ref{figwq}). 

\begin{figure}[ht]
\includegraphics[clip,angle=0,width=8cm]{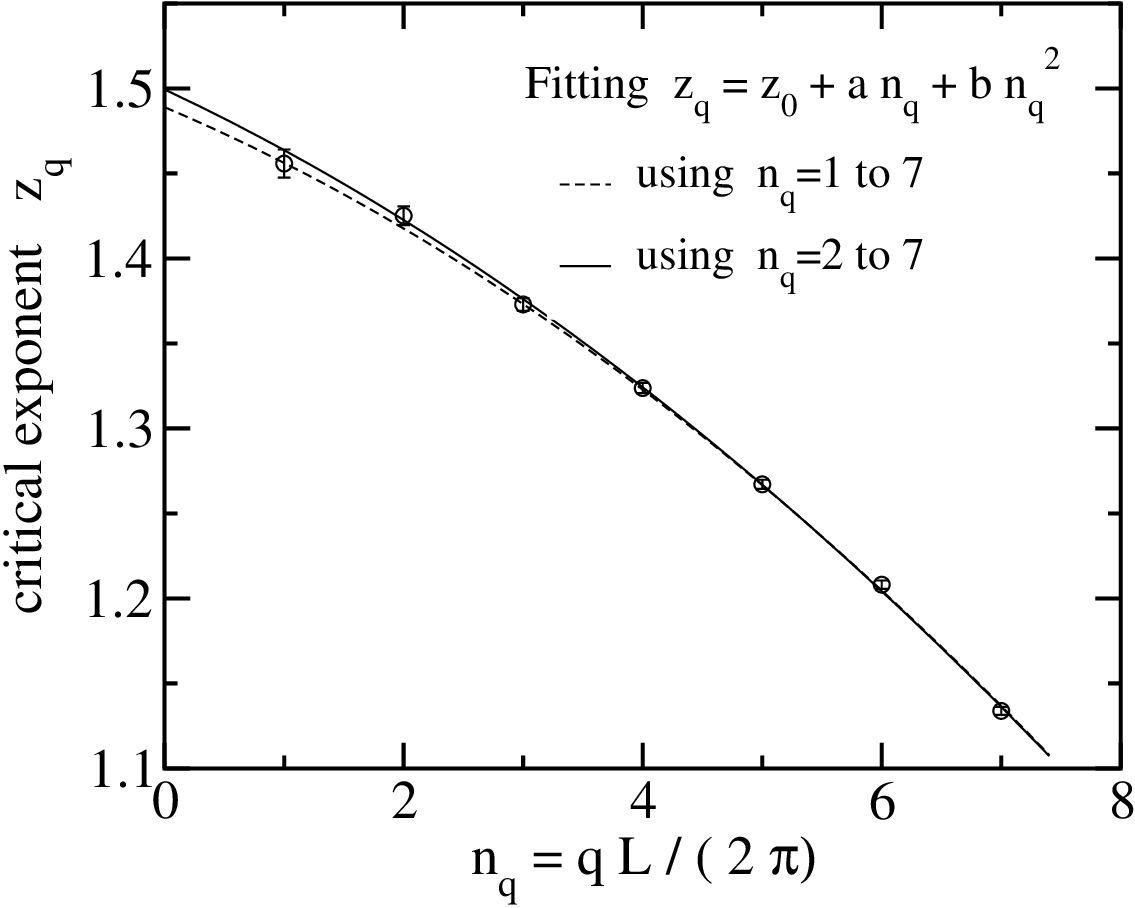}
\caption{\label{figz}Estimate of the dynamic critical exponent for 
different $n_q$. Analysis done with $L=30,36,48,60$, and $72$.}
\end{figure}

A better approach to determine $z$ is to use 
the dynamic finite-size scaling theory outlined above. Using this method,
we obtained $z$ for different values of $n_q$, with no resolution function.
Such estimates are denoted as $z_q$ and are shown in Fig.\ref{figz}.
In our previous work\cite{sd1}, we estimated $z$ as the average value
obtained using $n_q=1$ and $n_q=2$, with a maximum lattice size of $L=60$.  
We now have larger $L$ and better statistics, and hence the present
$n_q=1$ and $n_q=2$ correspond to $q$-values that are closer to the 
asymptotic critical region.  
Nevertheless, Fig.\ref{figz} shows that these values
of $q$ are not yet in the asymptotic critical region, and a better
estimate of $z$ is given by $z_0$, obtained by extrapolating $z_q$ to the 
limit $q\to 0$. We fitted $z_q$ with the function $z_q=z_0 + a\:n_q+b\:n_q^2$, 
where $z_0$, $a$, and $b$ are fitting parameters. Using $n_q=1,2,...,7$ in
the fitting (dashed line in Fig.\ref{figz}) we find $z_0=1.48\pm 0.02$ and
excluding $n_q=1$, due to its large finite-size dependence,
(solid line in Fig.\ref{figz}) we find $z_0=1.50\pm 0.02$. Both estimates
agree with the theoretical prediction of $z=1.5$.
Dynamic finite-size scaling theory with a small resolution function
was used to determine $z_q$ iteratively. The 
results thus obtained are within a one-$\sigma$ error bar of the 
respective $\delta_{\omega}=0$ estimates.

\begin{figure}[ht]
\includegraphics[clip,angle=0,width=8cm]{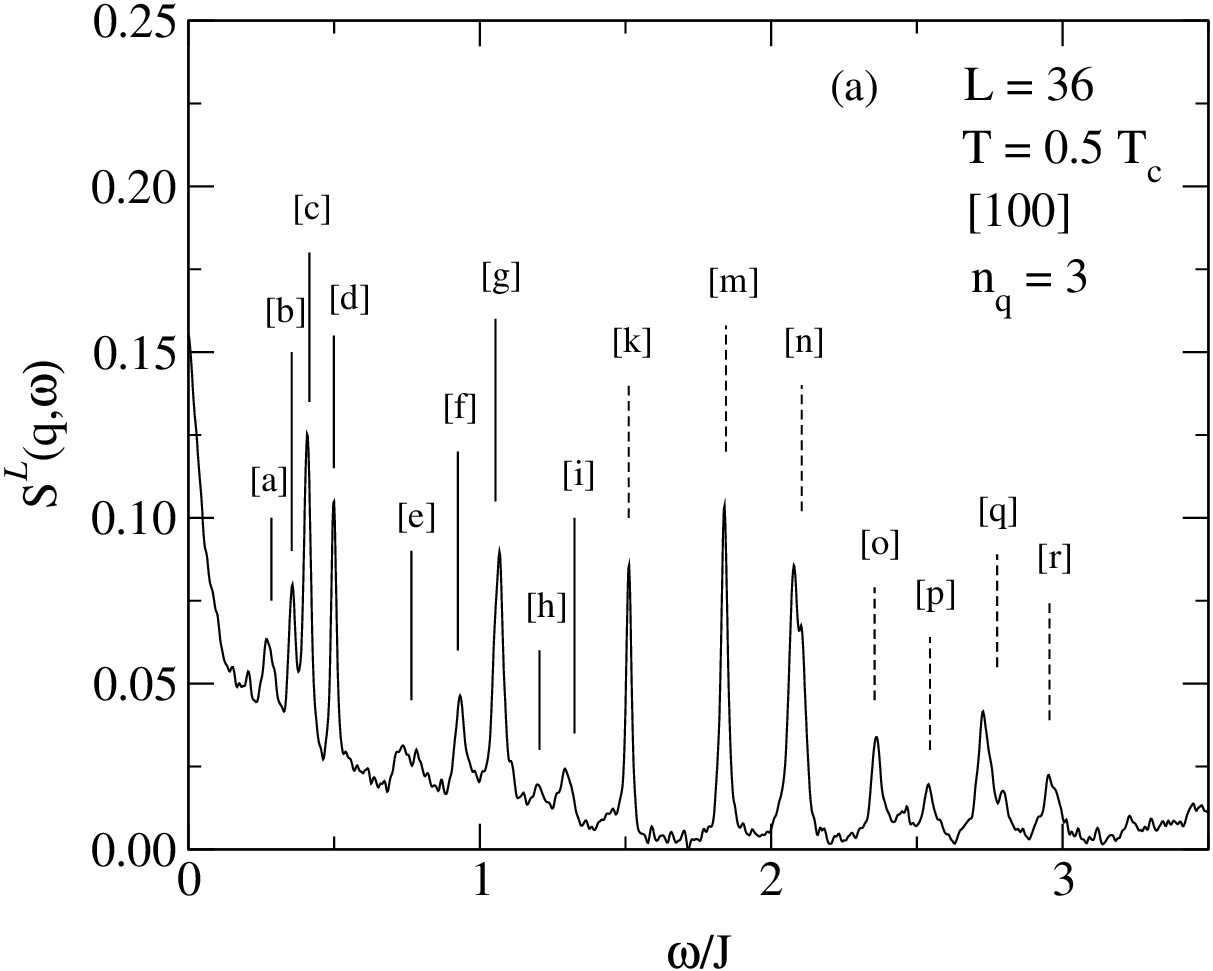}
\includegraphics[clip,angle=0,width=8cm]{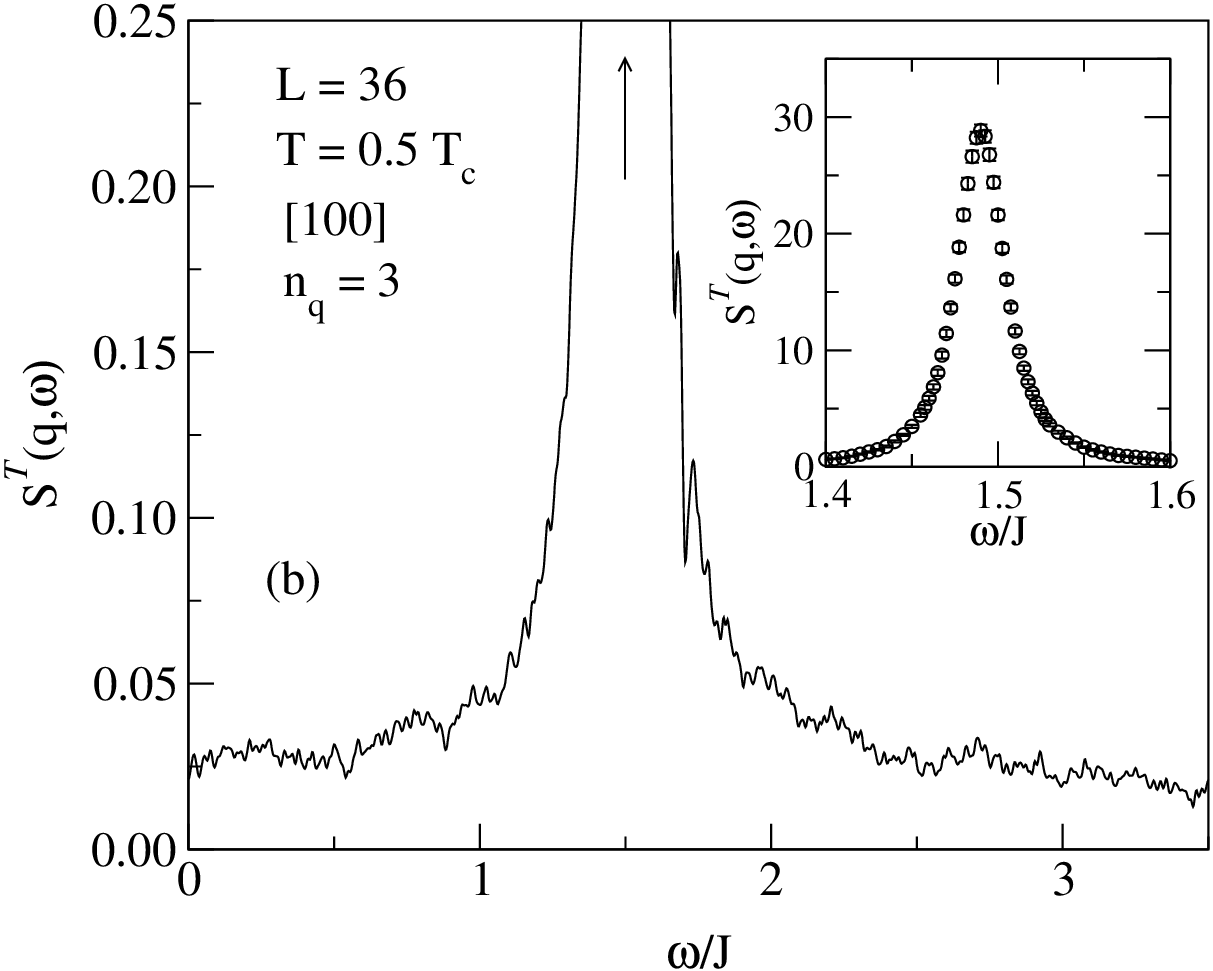}
\caption{\label{figsqwTp5}(a)Longitudinal and (b) transverse components of 
$S(q,\omega)$, with respect to the staggered magnetization, with 
$\delta_{\omega}=0.005$ at $T=0.5T_c$. Beside a
central peak, the longitudinal component has two-spin-wave
subtraction (indicated by solid lines) and addition peaks (dashed lines), 
corresponding to $18{\bf q}_1/\pi$ equal to [a] (2,2,0), [b] (1,1,1), 
[c] (1,1,0), [d] (1,0,0), [e] (3,2,0), [f] (3,1,1), [g] (3,1,0), 
[h] (4,1,1), [i] (2,1,1), [k] (1,0,0), [m] (1,1,0), [n] (1,1,1,) and (3,1,0), 
[o] (3,1,1), [p] (2,2,0), [q] (2,1,1) and (3,2,0), [r] (3,2,1) and 
(4,1,1).}
\end{figure}

The longitudinal and transverse components of the dynamic structure factor
with respect to the staggered magnetization
are shown in Figs.\ref{figsqwTp5}a and \ref{figsqwTp5}b, respectively, 
for $L=36$, $T=0.5T_c$ and ${\bf q}$ in the [100] direction with $n_q=3$. 
While the transverse component has a pronounced
single spin-wave excitation at $\omega/J\approx 1.49$ and intensity 
$\sim 30$, as shown in the inset in Fig.\ref{figsqwTp5}b, the structures 
on $S^L({\bf q},\omega)$ have much smaller
amplitudes and comprise a central peak, and a series of 
two-spin-wave subtraction (a-i) and addition (k-r) peaks. 
Denoting the momentum and frequency of two single spin
waves as $({\bf q}_1,\omega_1)$ and $({\bf q}_2,\omega_2)$, the excitations
resulting from their addition and subtraction have 
frequencies $\omega_+=\omega_1+\omega_2$
and  $\omega_-=|\omega_1-\omega_2|$, respectively, and momentum 
${\bf q}={\bf q}_1+{\bf q}_2$. For odd values of $n_q$ there are no two
spin-wave peaks at $\omega=0$ so the central peak seen in 
$S^L({\bf q},\omega)$ is presumably due to spin diffusion.

\begin{figure}[ht]
\includegraphics[clip,angle=0,width=8cm]{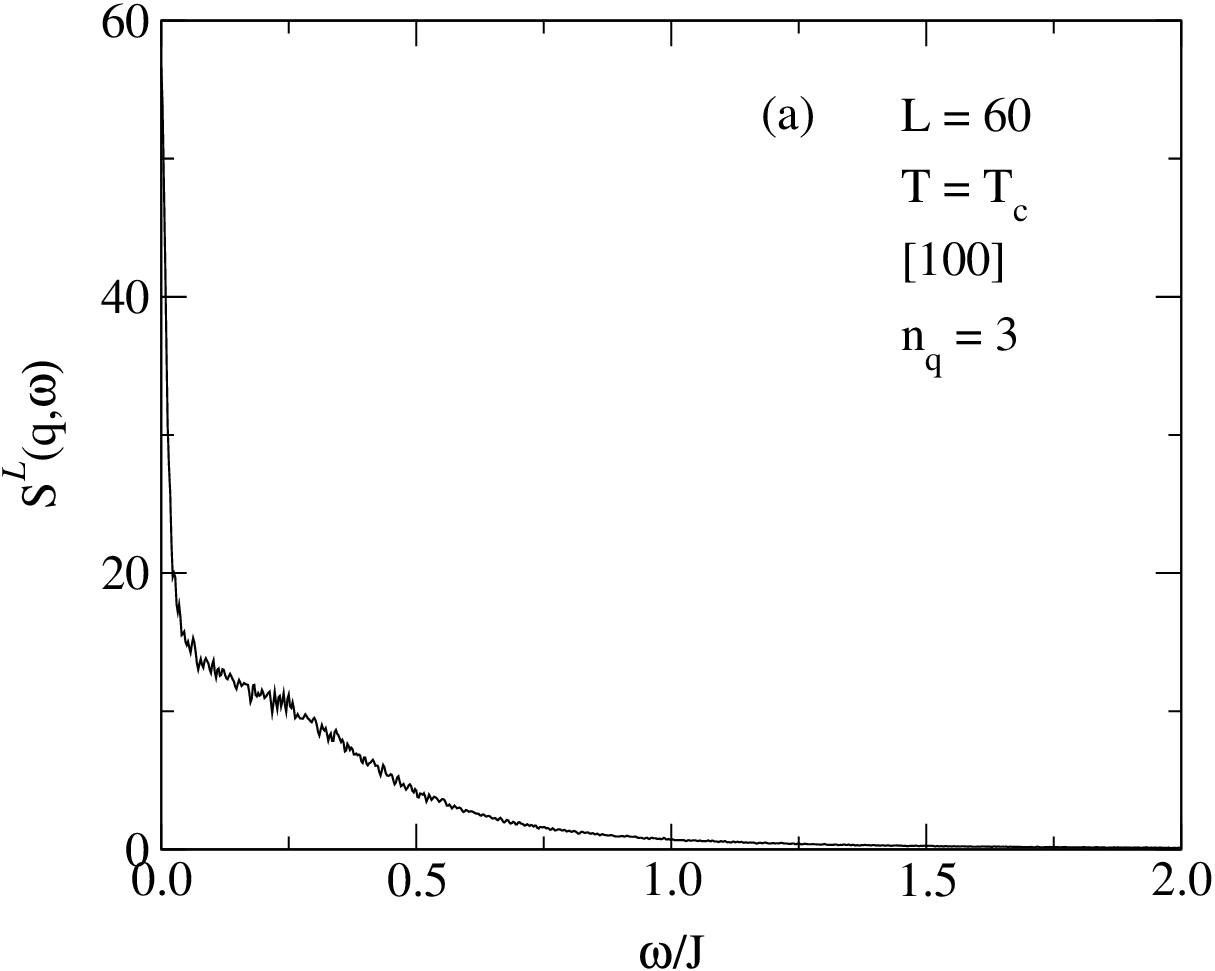}
\includegraphics[clip,angle=0,width=8cm]{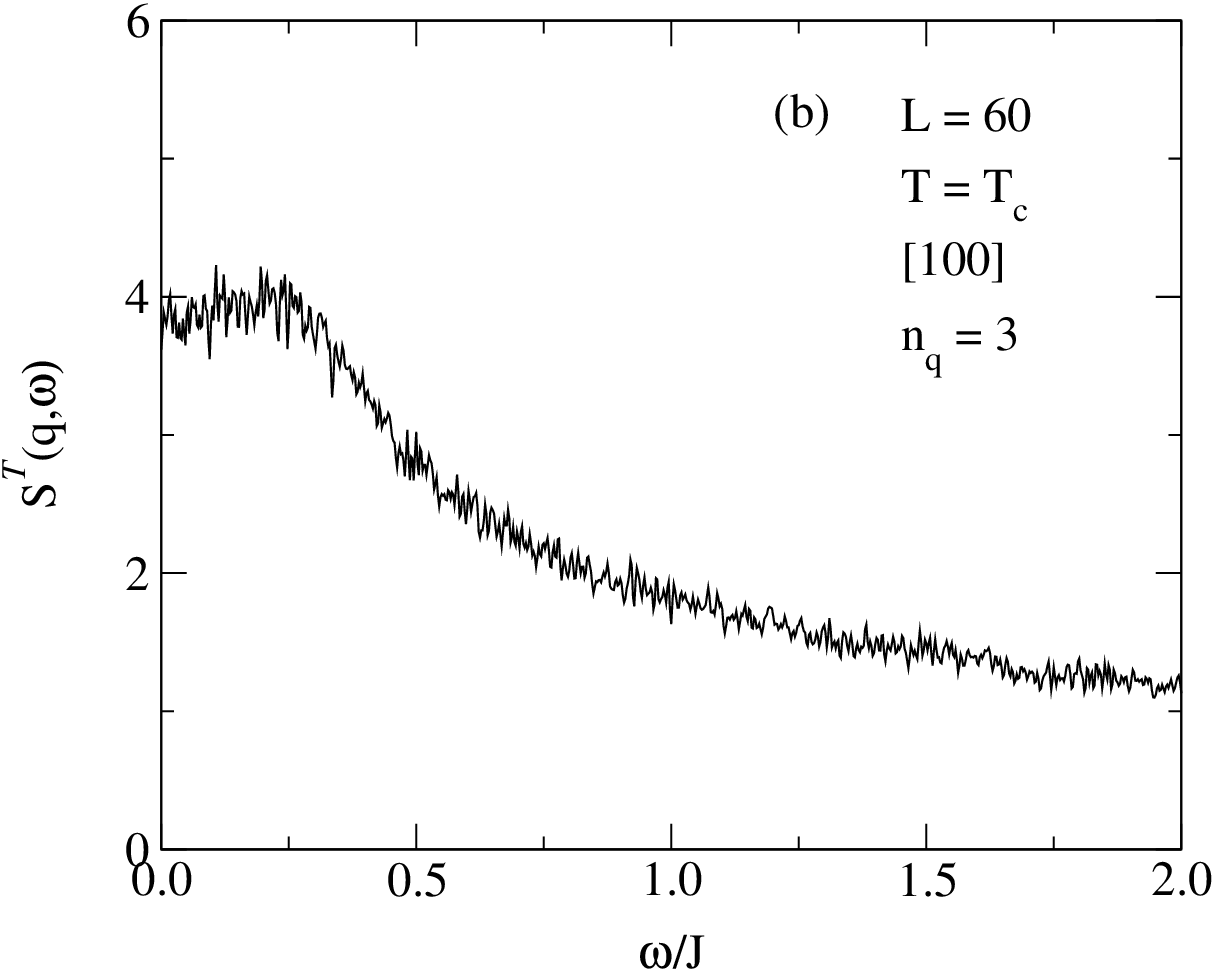}
\caption{\label{figsqwTc}(a)Longitudinal and (b) transverse components of 
$S(q,\omega)$, with respect to the staggered magnetization at $T_c$.}
\end{figure}

At $T_c$, $S^L({\bf q},\omega)$ (see Fig.\ref{figsqwTc}a) seems to be 
predominantly diffusive with a central peak that is much more intense than 
the spin-wave peak in $S^T({\bf q},\omega)$ (Fig.\ref{figsqwTc}b). 
These data provide clear evidence that the central peak in the average 
$S({\bf q},\omega)$ at $T_c$, seen in experiment and previous simulations,
but not present in renormalization-group and mode-coupling theories, appears
only in the longitudinal component. 
Two spin-wave excitations at $T_c$ cannot be resolved.

\begin{figure}[ht]
\includegraphics[clip,angle=0,width=8cm]{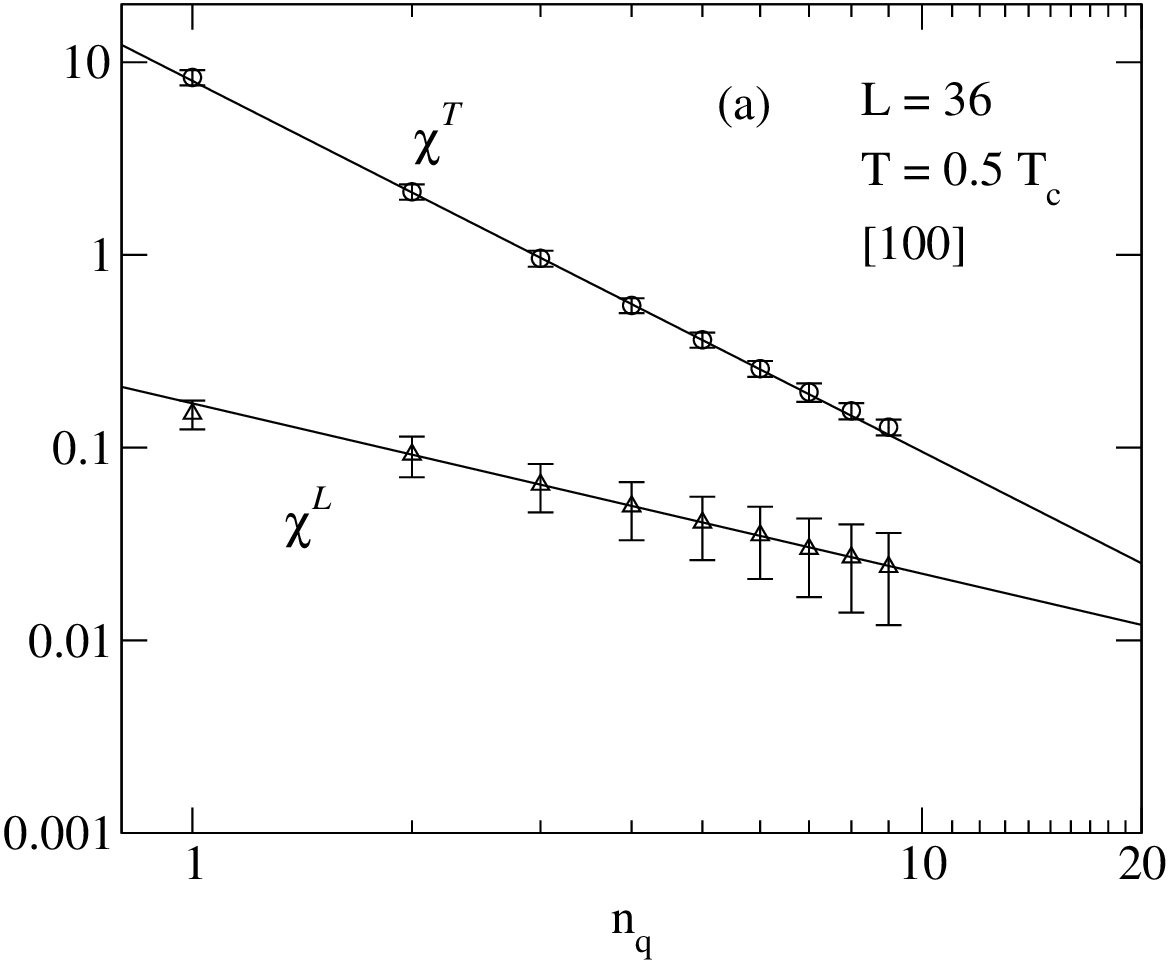}
\includegraphics[clip,angle=0,width=8cm]{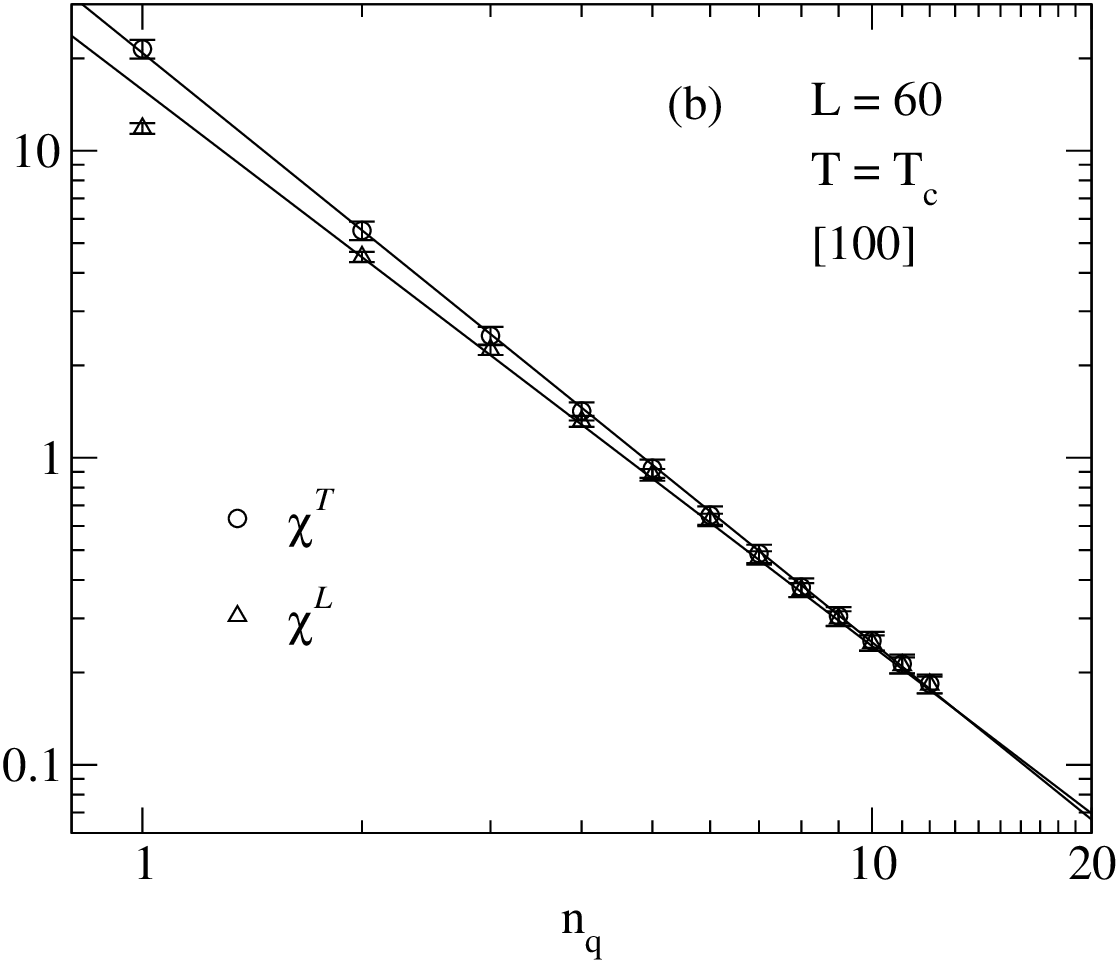}
\caption{\label{figxi}Log-log plot of the longitudinal ($\triangle$) and
 transverse ($\circ$) integrated intensity as a function of $n_q$, at 
(a) $T=0.5T_c$ and (b) $T=T_c$. The solid lines are linear fittings using 
$n_q=2$ to $n_q=6$ for (a), and to $n_q=10$ for (b).}
\end{figure}

Figure \ref{figxi} shows log-log plots of ${\chi}^L(q)$ and ${\chi}^T(q)$, the 
integrated intensities of $S^L(q,\omega)$ and $S^T(q,\omega)$,
respectively, as a function of momentum in the [100] direction. 
The momentum dependence of the integrated intensity has the form 
$q^{-{\rm x}}$.
At $T=0.5T_c$ (Fig.\ref{figxi}a), a linear fitting in the log-log plane
of ${\chi}^L(q)$ and ${\chi}^T(q)$ versus $n_q$ for $L=36$
using $n_q=1$ to $6$ gives $\chi^L\sim 1/q^{0.80\pm 0.16}$ and 
$\chi^T\sim 1/q^{1.94\pm 0.06}$, whereas if the $n_q=1$ point is dropped
 the linear fitting yields (solid lines) 
 $\chi^L\sim 1/q^{0.88\pm 0.35}$ and $\chi^T\sim 1/q^{1.92\pm 0.10}$. These 
results are in agreement with both renormalization-group theory 
prediction\cite{mazenkoTb} and experimental results\cite{coldea}.
We have also tried to fit $\chi^L(q)$ with the Ornstein-Zernike mean-field 
formula\cite{collinsbook} $\chi^L(q)=\chi^L(0)\;\kappa^2/(q^2+\kappa^2)$, 
using $\chi^L(0)$ and $\kappa$ as
fitting parameters; however, this expression did not yield a good fitting.
\begin{figure}[ht]
\includegraphics[clip,angle=0,width=8cm]{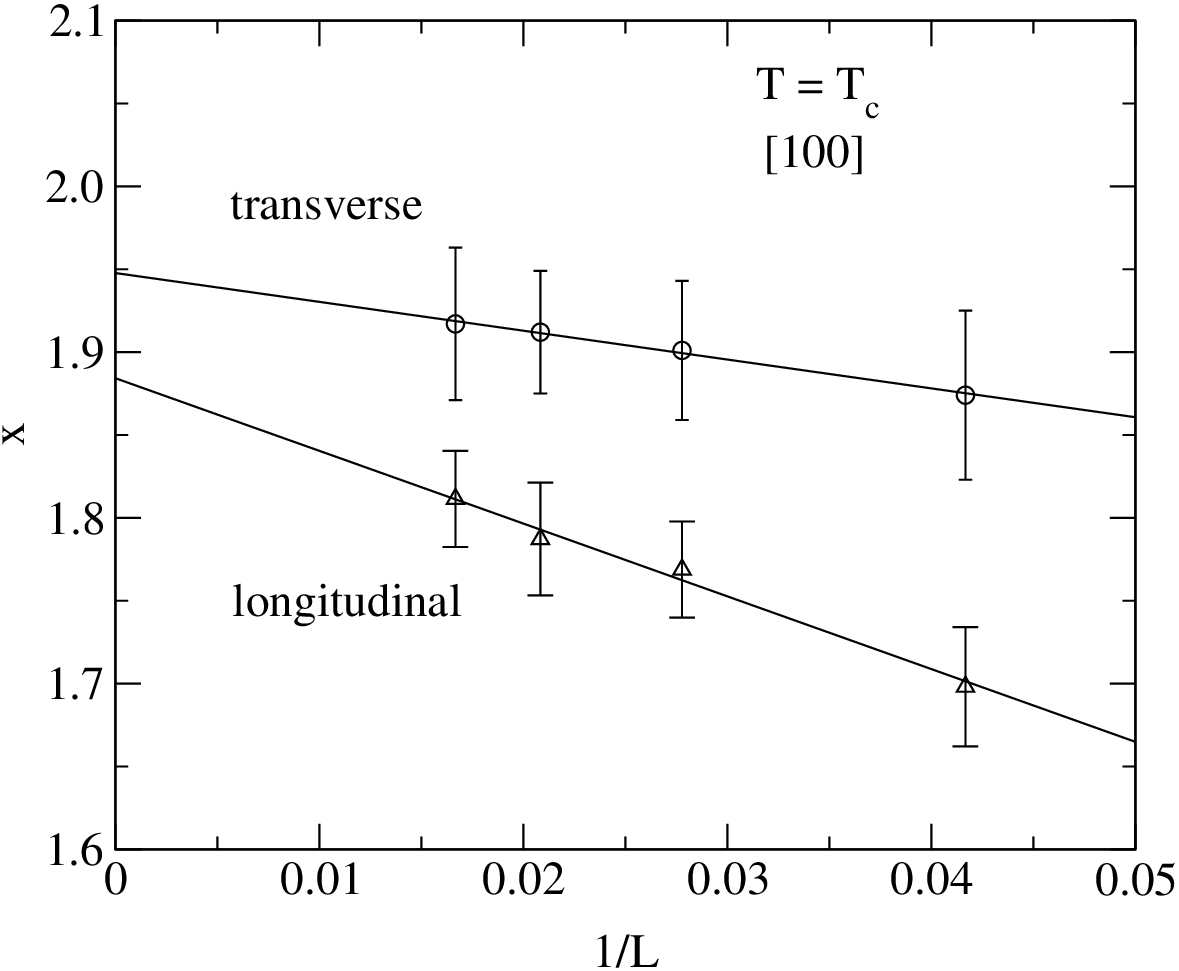}
\caption{\label{figxiexp}Low-$q$ divergence exponents of 
$\chi^L$ ($\triangle$) and $\chi^T$ ($\circ$) at $T_c$, 
as a function of the inverse lattice linear size.
The solid lines are linear fittings using data for $L=24,36,48,60$.}
\end{figure}
At $T_c$ (Fig.\ref{figxi}b), our data for $L=60$ indicate that 
$\chi^L\sim 1/q^{1.81\pm 0.03}$ 
and $\chi^T\sim 1/q^{1.92\pm 0.04}$, where $n_q=2$ to $10$ have been used in
the fitting. Finite-size effects on the low-$q$ divergence exponents
of $\chi^L$ and $\chi^T$ at $T_c$ are shown in Fig.\ref{figxiexp}. 
A linear fitting of these exponents as a
function of $1/L$ yields $\chi^L\sim 1/q^{1.88\pm 0.05}$ and 
$\chi^T\sim 1/q^{1.95\pm 0.07}$ for 
the thermodynamic limit where $L=\infty$. The dynamic scaling prediction
for the static susceptibility at $T_c$ is\cite{halhoh69,RitFis72} 
$\chi=1/q^{2-\eta}$, where 
for the purpose of this comparison we can use\cite{eta_ref} 
$\eta\approx 0.04\pm 0.01$. We see that while $\chi^T$ is consistent with
the dynamic scaling prediction, our large error bars do not exclude 
the mean-field behavior of $\chi\sim 1/q^2$. Our estimate for 
the divergence of $\chi^L$ at small $q$ is slightly less rapid than predicted,
but still consistent with it within a two-$\sigma$ error bar. 

\begin{figure}[ht]
\includegraphics[clip,angle=0,width=8cm]{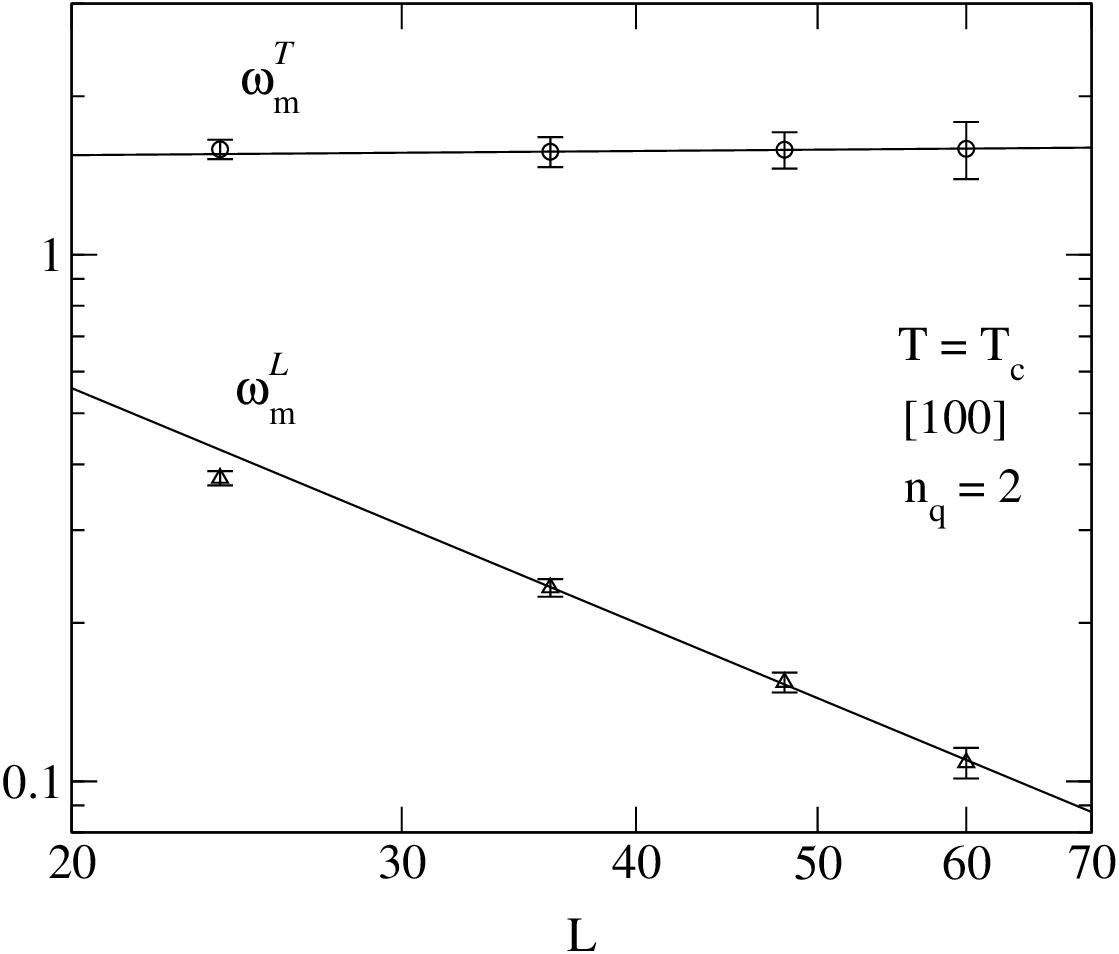}
\caption{\label{figwmLLT}Log-log plot of the longitudinal ($\triangle$) and
 transverse ($\circ$) components of the characteristic frequency as a 
function of $L$ with $\delta_{\omega}=0$.} 
\end{figure}

Figure \ref{figwmLLT} shows a log-log plot of the longitudinal and transverse
components of the characteristic frequency as a function of $L$, for $n_q=2$  
and $\delta_{\omega}=0$ at $T_c$. These components are denoted as 
$\omega_m^L$ and $\omega_m^T$, respectively, and according to dynamic 
finite-size scaling we have $\omega_m^L=L^{-z^L}\Omega^L(qL)$ and 
$\omega_m^T=L^{-z^T}\Omega^T(qL)$. For $n_q=2$ (see Fig.\ref{figwmLLT}) 
we obtain $z^L=1.48\pm 0.14$ and $z^T=-0.03\pm 0.25$, where data for 
$L=36,48$, and $60$ have been included in the analysis.
Our data show that the dynamic critical exponent for the transverse component 
is consistent with $z=0$, indicating that this component is 
not critical. In contrast, the longitudinal component is critical, with 
$z\approx 1.5$. 

Separating $S({\bf q},\omega)$ into longitudinal and transverse components
with respect to the uniform magnetization we see that the 
longitudinal component has a central peak, a pronounced spin-wave peak 
and less intense two-spin-wave peaks. In contrast, each such peak in the 
transverse component is split into two peaks, with frequencies 
$\omega^L\pm \Delta\omega$, where $\omega^L$ is the frequency of the 
corresponding peak in the longitudinal component of $S({\bf q},\omega)$.
The shift $\Delta\omega$ in the peak frequencies corresponds to the
frequency of oscillation of the staggered magnetization.

\section{Conclusions}

We have used Monte Carlo and spin dynamics simulations to 
study the dynamic behavior of the isotropic Heisenberg antiferromagnet 
on the simple cubic lattice.  
When we use the same range of momentum $q$ as probed by 
experiment\cite{coldea}, the dynamic critical exponent obtained\cite{sd1} 
is in good agreement with its experimental value, which is slightly lower
than theoretical predictions. In our present work we have used a larger 
lattice size, and thus smaller values of $q$, in addition to obtaining better 
statistics and extrapolating finite $q$ results to the limit $q=0$. 
This allowed us to study systematic changes as we approach the asymptotic 
critical region and our improved estimate ({\it i.e.} with systematic 
errors largely eliminated) is $z=1.49\pm 0.03$, in good agreement with
the renormalization group theory and dynamic scaling predictions. 
Presumably the values of $q$ used in the experiment\cite{coldea} and in our
previous simulations were not in the true asymptotic critical
region, resulting in a slightly lower estimate of $z$.
This should serve as a warning for future simulational and experimental
probes of dynamic critical behavior.

Longitudinal and transverse components of $S({\bf q},\omega)$ with
respect to the staggered magnetization are investigated separately. 
This required rotation of the frame of reference after each integration
step because the staggered magnetization is not a conserved quantity.
Below $T_c$, the transverse component 
$S^T({\bf q},\omega)$ has a pronounced spin-wave peak, whereas the 
longitudinal component $S^L({\bf q},\omega)$ is dominated by two-spin-wave
addition and subtraction peaks, and a central peak presumably due to spin 
diffusion. These results are consistent with theory\cite{mazenkoTb} 
and experiment\cite{cox}, both of which 
have shown that the transverse spin fluctuations are propagating, dominated
by spin-waves, whereas the quasielastic peak is due to longitudinal
fluctuations. 
At $T_c$, $S^L({\bf q},\omega)$ has a central peak that is much more 
intense than the spin-wave peak in $S^T({\bf q},\omega)$ and no central peak
was seen in $S^T({\bf q},\omega)$. Explaining the appearance of a central peak
in $S^L({\bf q},\omega)$ at $T_c$ remains a challenge for theory. We have 
also seen that while $S^T({\bf q},\omega)$ is not critical, 
$S^L({\bf q},\omega)$ is critical and it has a dynamic critical exponent 
$z\approx 1.5$.

These findings further support our earlier conclusion that a simple, 
nearest-neighbor, isotropic Heisenberg model describes the behavior of
RbMnF$_3$ quite well. The only limitations in the agreement appear to be
at $T_c$, and even there all qualitative features and dynamic exponent are
faithfully reproduced.

Below $T_c$ our results for the longitudinal and transverse components of the
integrated intensities of $S({\bf q},\omega)$ are consistent with 
renormalization-group theory predictions, and not with the mean-field one.
In contrast, at $T_c$, while the integrated intensities are consistent with 
the renormalization-group theory prediction, the large error bars 
do not allow us to exclude the mean-field behavior.

\section{Acknowledgments}
Fruitful discussions with A. Cuccoli are gratefully acknowledged.
This research was partially supported by NSF grant DMR-0094422.
Simulations were performed on the Cray T90 at SDSC and on the IBM SP 
at the U. of Michigan.

\end{document}